\documentclass[twocolumn]{aastex62}
\pdfoutput=1
\usepackage{graphicx}

\usepackage[normalem]{ulem}
\usepackage{amsmath}

\usepackage{dcolumn}
\usepackage{bm}
\usepackage{natbib}

\renewcommand{\arraystretch}{1.5}

\submitjournal{ApJL}

\begin{document}

\title{Limits on the number of spacetime dimensions from GW170817}

\author{Kris Pardo}
\email{kpardo@astro.princeton.edu}
\affiliation{Department of Astrophysical Sciences, Princeton University,\\
Princeton, NJ 08544, USA}

\author{Maya Fishbach}
\affiliation{Department of Astronomy \& Astrophysics, University of Chicago,\\
Chicago, Illinois 60637, USA}

\author{Daniel E. Holz}
\affiliation{Department of Astronomy \& Astrophysics, University of Chicago,\\
Chicago, Illinois 60637, USA}
\affiliation{Enrico Fermi Institute, Department of Physics, and Kavli Institute
for Cosmological Physics, University of Chicago,\\
Chicago, Illinois 60637, USA
}
\affiliation{Kavli Institute for Particle Astrophysics \& Cosmology and Physics Department, Stanford University,\\ Stanford, CA 94305}

\author{David N. Spergel}
\affiliation{Department of Astrophysical Sciences, Princeton University,\\
Princeton, NJ 08544, USA}
\affiliation{Center for Computational Astrophysics, Flatiron Institute,\\ New York, NY 10003, USA}

\date{\today}

\begin{abstract}
The observation of GW170817 in both gravitational and electromagnetic waves provides a number of unique tests of general relativity. One question we can answer with this event is: Do large-wavelength gravitational waves and short-frequency photons experience the same number of spacetime dimensions? In models that include additional non-compact spacetime dimensions, as the gravitational waves propagate, they ``leak'' into the extra dimensions, leading to a reduction in the amplitude of the observed gravitational waves, and a commensurate systematic error in the inferred distance to the gravitational wave source. Electromagnetic waves would remain unaffected. We compare the inferred distance to GW170817 from the observation of gravitational waves, $d_L^\mathrm{GW}$, with the inferred distance to the electromagnetic counterpart NGC 4993, $d_L^\mathrm{EM}$. We constrain $d_L^\mathrm{GW} = (d_L^\mathrm{EM}/\mathrm{Mpc})^\mathrm{\gamma}$ with $\gamma = 1.01^{+0.04}_{-0.05}$ (for the SHoES value of $H_0$) or $\gamma = 0.99^{+0.03}_{-0.05}$ (for the Planck value of $H_0$), where all values are MAP and minimal 68\% credible intervals.
These constraints imply that gravitational waves propagate in $D=3+1$ spacetime dimensions, as expected in general relativity. In particular, we find that $D = 4.02^{+0.07}_{-0.10}$ (SHoES)  and $D = 3.98^{+0.07}_{-0.09}$ (Planck).
Furthermore, we place limits on the screening scale for theories with $D>4$ spacetime dimensions, finding that the screening scale must be greater than $\sim 20$\,Mpc. We also place a lower limit on the lifetime of the graviton of $t > 4.50 \times 10^8$\,yr.
\end{abstract}

\section{Introduction}
Gravitational wave (GW) events with electromagnetic (EM) counterparts are powerful tests of modified gravity theories. Importantly, such joint observations are sensitive to differences between the propagation of GW and EM waves through spacetime. The recent detection of the first multi-messenger GW system, GW170817 \citep{LIGO2017}, allows us to constrain modified gravity in this way for the first time.

From the time delay between the electromagnetic and GW signals, powerful limits can be placed on the speed of GW propagation \citep{LIGOboth}. Many papers have already discussed how this constrains specific modified gravity theories \citep[e.g.,][]{Lombriser2016, Lombriser2017,EzquiagaZumalac2017,Baker2017,Creminelli2017,Visinelli2017,Sakstein2017,Nersisyan2018}. 

The independent distance measures of the GW source and its EM counterpart can also place constraints on the damping of GWs. Since GWs are standard sirens, we can directly extract the luminosity distance to the GW source \citep{Schutz1986,Holz2005,Dalal2006,Nissanke2010,Nissanke2013,2017arXiv171206531C}. In addition, we can make an independent measurement of the distance to the source by measuring the redshift of the EM counterpart and using our knowledge of cosmology (in particular, the Hubble constant, since GW170817 is at low redshift) to convert the observed redshift into a luminosity distance. By comparing these two distances, we can place limits on the damping of GWs. A number of authors have discussed the power of gravitational waves sources to place these sorts of constraints \citep{Nishizawa2017,Arai2017,Belgacem2017,Amendola2017,2018arXiv180101503L}; in what follows we focus on general constraints provided by the recent observations of GW170817 and its associated EM counterpart.

In this paper we constrain GW damping by considering modifications to the signal's attenuation with luminosity distance. According to GR, the GW amplitude decreases inversely with luminosity distance. However, extra-dimensional theories of gravity with non-compact extra dimensions generally predict a deviation from this relationship. Comparing the luminosity distance of GW170817 extracted under the assumption of GR to the EM-measured distance to its host galaxy, NGC 4993, we find stringent constraints on theories with gravitational leakage. We use these limits to set bounds on the number of additional non-compact spacetime dimensions and characterize properties of the modifications, such as the screening scale and the lifetime of the graviton. Section 2 describes the waveforms that we consider and gives a qualitative description of our analysis. Section 3 describes our methods. Section 4 gives our results and explores other applications.

\section{Gravitational leakage and gravitational waves}
In this section we summarize the effects of gravitational leakage on the GW waveform and its relation to higher-dimensional theories. We also give a qualitative introduction to how GW170817 constrains gravitational leakage. This section relies heavily on the work of \cite{Deffayet2007}.

In GR the strain goes as:
\begin{equation}\label{eqn:grwaveform}
h_{\rm{GR}} \propto \frac{1}{d_L},
\end{equation}
where $d_L$ is the luminosity distance of the GW source. For a higher-dimensional theory where there is some leakage of gravity we would expect, due to flux conservation, damping of the wave in the form of a power-law\citep{Deffayet2007}:
\begin{equation}\label{eqn:dampdwaveform}
h \propto \frac{1}{d_L^{\gamma}},
\end{equation}
where $\gamma$ is related to the number of dimensions, $D$, by:
\begin{equation}\label{eqn:gammadim}
\gamma = \frac{D-2}{2}.
\end{equation}
More generally, we may consider theories that have an associated screening scale, $R_c$. These theories behave like GR below this scale, but exhibit gravitational leakage above $R_c$. In such theories the GW strain scales as \citep{Deffayet2007}:
\begin{equation}\label{eqn:defmenwaveform}
h \propto \frac{1}{d_L\left[ 1 + \left(\frac{d_L}{R_c}\right)^{n(D-4)/2} \right]^{1/n}},
\end{equation}
where $n$ gives the transition steepness. This waveform reduces to Equation~\ref{eqn:dampdwaveform} for $d_L \gg R_c$.

Finally, we consider theories in which the graviton has a decay channel. In this case, the amplitude of the GW would scale as:
\begin{equation}
\label{eqn:decay}
h \propto \frac{\exp \left[-d_L/R_g \right]}{d_L},
\end{equation}
where $R_g$ is the `decay-length' (i.e. the distance a graviton travels during its average lifetime).

If we assume that, outside of these overall damping factors, the waveforms remain unchanged from the predicted GR form, then the gravitational leakage would simply result in a measured $d_L$ greater than the true $d_L$ for the source (i.e. the GW would appear to have come from farther away because it would have a smaller amplitude in the detectors). An event only measured in GWs would not allow us to distinguish the measured $d_L$ from the true value. However, GW170817 was also detected electromagnetically; thus, we have an independent measurement of the luminosity distance for this source. By comparing the measured GW distance and the measured EM distance, we can constrain the gravitational leakage parameter $\gamma$ (defined in Equation~\ref{eqn:dampdwaveform}) and therefore place limits on the number of spacetime dimensions, the screening scale, or the lifetime of the graviton.  
In this we implicitly assume that the luminosity distance inferred from EM observations is the true luminosity distance: $d_L^{\rm EM}=d_L$; in practice, our approach quantifies the difference between the EM and GW distance estimates, and is insensitive to the true value of $d_L$.

Since the GW170817 standard siren measurement of the Hubble constant is consistent with expectations \citep{H0paper}, this implies that, for reasonable assumed values of the Hubble constant, the inferred GW and EM distances are similarly consistent. We therefore expect that general relativity provides an excellent description, and we would not expect strong evidence for gravitational leakage and extra dimensions. In what follows we quantify this expectation.

\section{Method}

In order to measure gravitational leakage, we compare the EM luminosity distance to the source, $d_L^\mathrm{EM}$, with the GW luminosity distance, $d_L^\mathrm{GW}$, extracted from the waveform under the assumption that GR is the correct theory of gravity. To find the EM luminosity distance to the source, we use Hubble's law to relate the host galaxy's ``Hubble velocity", $v_H$, to its luminosity distance. In the nearby universe, this relationship can be approximated by: 
\begin{equation}
v_H = H_0 d_L^\mathrm{EM}.
\end{equation}
The Hubble velocity is the recessional velocity that the galaxy would have if it was stationary with respect to the Hubble flow. To find the Hubble velocity of the host galaxy NGC 4993, we follow \cite{H0paper} and correct the recessional velocity of the galaxy group to which NGC 4993 belongs, ESO-508, by its peculiar velocity. The EM observables are then the measured recessional velocity, $v_r$, of the group of galaxies to which NGC 4993 belongs, and the measured peculiar velocity, $\langle v_p \rangle$, in the neighborhood of NGC 4993. We denote the true peculiar velocity by $v_p$, so that the true recessional velocity is the sum of $v_H$ and $v_p$. We adopt the conservative uncertainty on $v_p$ from \cite{Guidorzi2017}, which sets the Hubble velocity to be $v_H = 3017 \pm 250$ km s$^{-1}$. Together with a prior measurement of the Hubble constant, the measured velocities, $v_r$ and $\langle v_p \rangle$, yield a measurement of the EM luminosity distance to the system.

Meanwhile, the GW data, $x_\mathrm{GW}$, gives the posterior probability of the GW luminosity distance, $d_L^\mathrm{GW}$, marginalized over all other waveform parameters, except the sky position, which is fixed to the position of the optical counterpart. We recover the GW distance posterior from the LIGO-Virgo Collaboration's publicly available $H_0$ posterior samples \citep{H0paper}. The $H_0$ posterior is given by marginalizing the joint probability of $H_0$, the GW distance posterior probability, $p(d_L^\mathrm{GW} \mid x_\mathrm{GW})$, and the velocities $v_H$ and $v_p$, over all parameters except $H_0$ (Eq.~9 of \cite{H0paper}). We recover the GW distance posterior (marginalized over inclination angles) from the $H_0$ posterior by deconvolving the $v_r$ and $v_p$ terms, which are given by Gaussians. We approximate the integral in Equation~9 of \cite{H0paper} by a Riemann sum. Then the term $p(x_\mathrm{GW} \mid d_L^\mathrm{GW})p(d_L^\mathrm{GW})$ is obtained by solving a system of linear equations.

\begin{figure}
\includegraphics[width=0.5\textwidth]{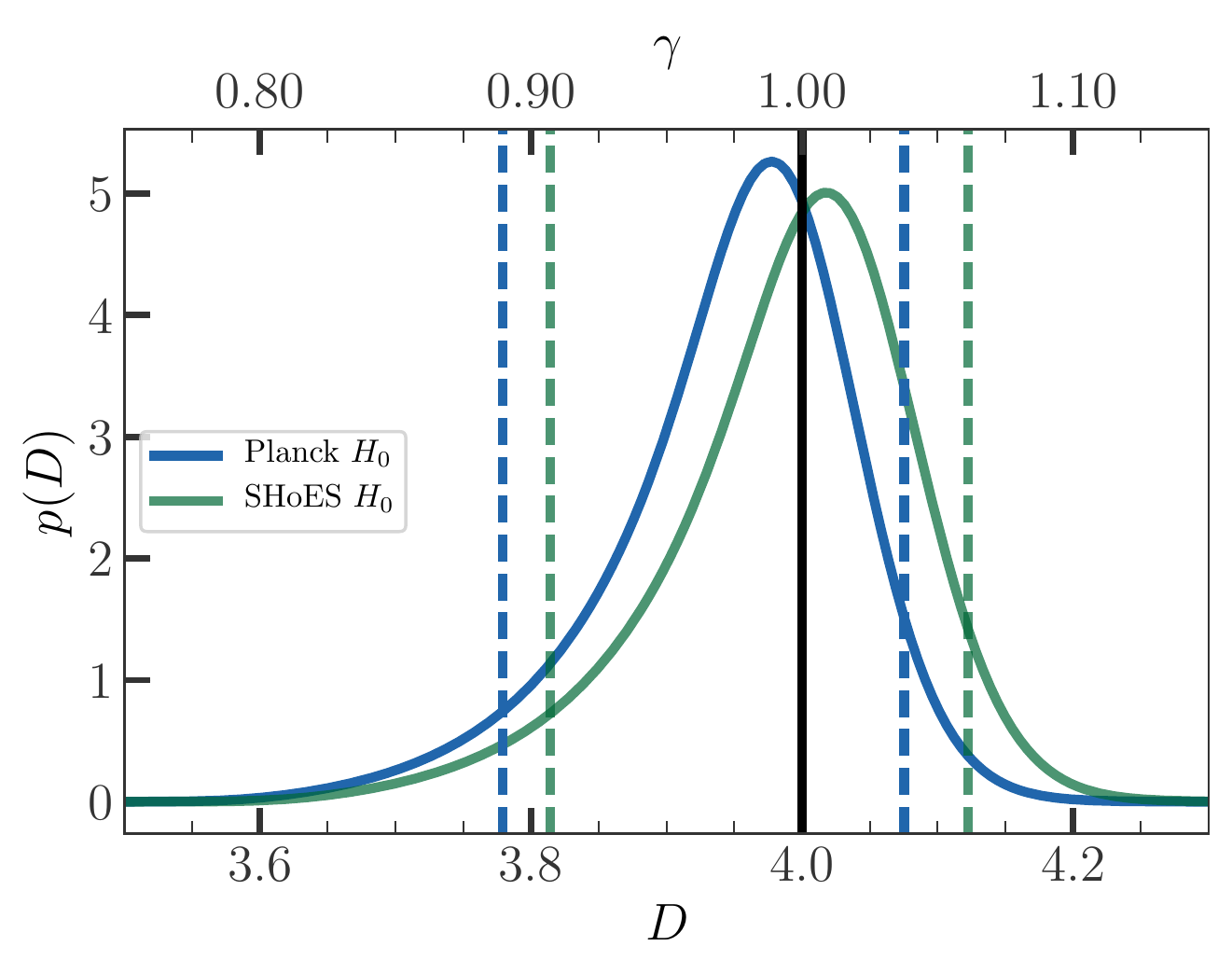}
\caption{\label{fig:gammaposterior} Posterior probability distribution for the number of spacetime dimensions, $D$, using the GW distance posterior to GW170817 and the measured Hubble velocity to its host galaxy, NGC 4993, assuming the $H_0$ measurements from \cite{Planck} (blue curve) and \cite{SHoES} (green curve). The dashed lines show the symmetric 90\% credible intervals.
The equivalent constraints on the damping factor, $\gamma$, are shown on the top axis.
GW170817 constrains $D$ to be very close to the GR value of $D=4$ spacetime dimensions, denoted by the solid black line.}
\end{figure}

We carry out a Bayesian analysis to infer the posterior of the gravitational leakage parameter, $\gamma$, and the number of spacetime dimensions, $D$, given the GW and EM measurements described above. The statistical framework is described in detail in the Appendix. 

\section{Results \& Discussion} 

\begin{table*}[bt!]
\centering
\begin{tabular}{||c|cc||}
\hline 
$H_0$ prior & $\gamma$ &  $D$\\
$\rm{km\ s^{-1}\ Mpc}^{-1}$&  &  \\ 
\hline
$H_0 = 73.24\pm 1.74$ \citep{SHoES}
& $1.01^{+0.04}_{-0.05}$ & $4.02^{+0.07}_{-0.10}$ \\
$H_0 = 67.74\pm 0.46$ \citep{Planck} &
$0.99^{+0.03}_{-0.05}$ & $3.98^{+0.07}_{-0.09}$\\
\hline
\end{tabular}
\caption{\label{tab-results1} Constraints on the damping parameter $\gamma$ and the number of dimensions $D$ assuming a waveform of the type Equation~\ref{eqn:dampdwaveform} from GW170817.} 
\end{table*}

The posterior for $D$ assuming a waveform with the scaling shown in Equations~\ref{eqn:dampdwaveform} and~\ref{eqn:gammadim} is given in Figure~\ref{fig:gammaposterior}. Since the results depend on the assumed $H_0$ prior, we compute the $D$ posterior for both the SHoES $H_0$ value \citep{SHoES} and the Planck $H_0$ value \citep{Planck}.
The maximum a posteriori (MAP) values and minimal 68\% credible interval values for $\gamma$ and $D$ are given in Table~\ref{tab-results1}. As can be seen, the results are completely consistent with GR.

We can also use these constraints to place limits on waveforms with a scaling given by Equation~\ref{eqn:defmenwaveform}. For the higher-dimensional theories that give rise to such waveforms, the $d_L^{\mathrm{GW}}$ measured under the assumption of GR will be greater than the true luminosity distance, $d_L^{\mathrm{EM}}$. Thus, while our posterior for $\gamma$ allows for both $\gamma > 1$ and $\gamma < 1$ (allowing for the relative damping of both the GW and EM signals), in the following analysis we restrict $\gamma > 1$. Using our joint posterior on $d_L^{\mathrm{GW}}$ and $d_L^\mathrm{EM} = (d_L^\mathrm{GW})^{1/\gamma}$ for GW170817, we can apply Equation~\ref{eqn:defmenwaveform} to constrain the screening radius, $R_c$:
\begin{equation}
\label{eqn:Rc}
R_c = \frac{d_L^\mathrm{EM}}{\left[ \left(\frac{d_L^\mathrm{GW}}{d_L^\mathrm{EM}}\right)^n - 1 \right]^{\frac{2}{n(D-4)}}}.
\end{equation}
Thus, given our posterior samples for $d_L^\mathrm{GW}$ and $\gamma$ (restricted to $\gamma > 1$), we can calculate the associated $R_c$ for a fixed transition steepness, $n$, and number of dimensions, $D$. Marginalizing over $H_0$ and $v_p$, this gives us a joint posterior on $R_c$ and $d_L^\mathrm{GW}$.

\begin{figure}
    \includegraphics[width=0.5\textwidth]{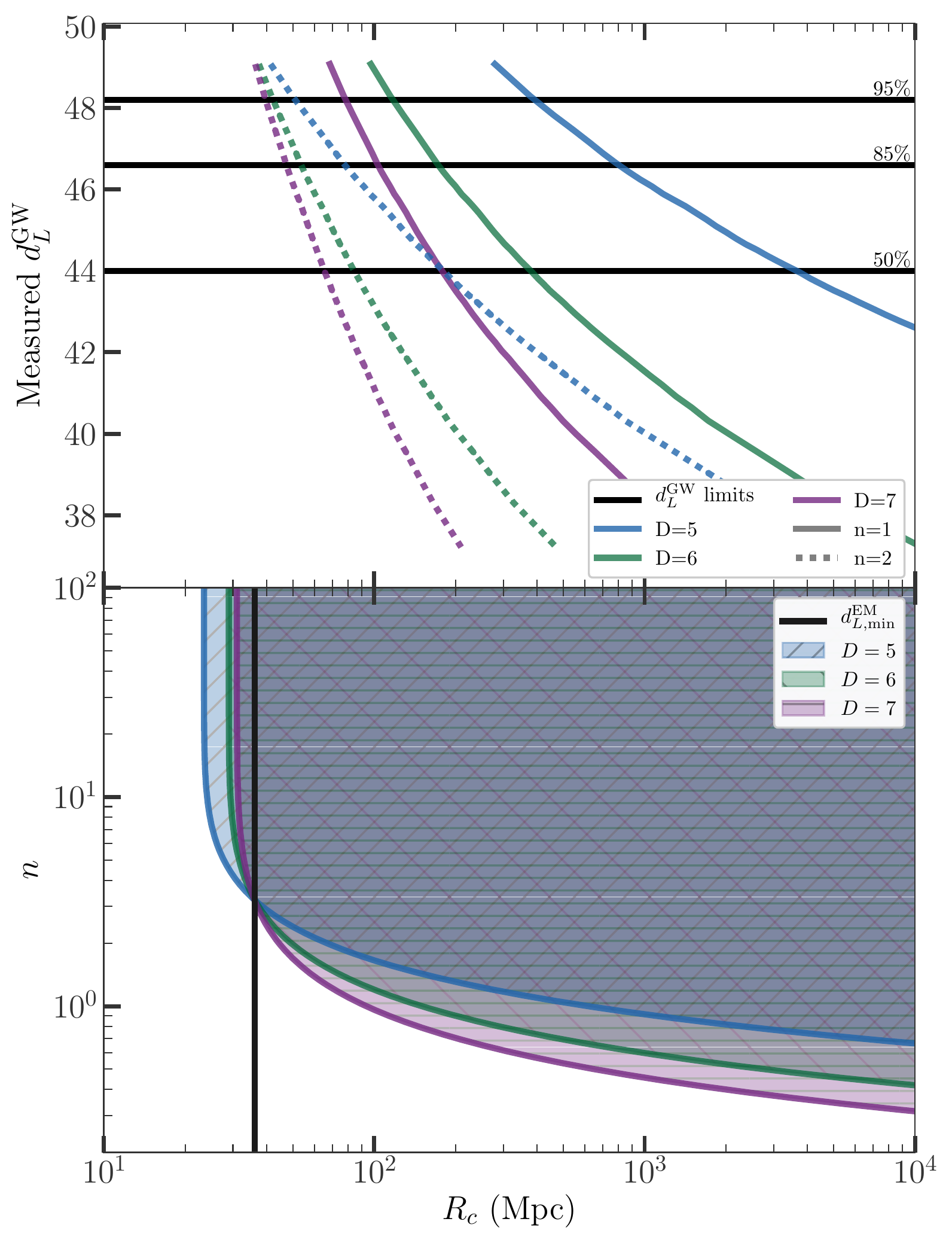}
    \caption{\label{fig:dgwvrc} \label{fig:twoparam} \textit{Top:} Measured luminosity distance from GWs, $d_L^{\mathrm{GW}}$ versus the gravitational screening scale, $R_c$, for a number of spacetime dimensions given by $D=5$ (blue), $D=6$ (green), and $D=7$ (purple). The solid lines assume a transition steepness of $n=1$ and the dotted lines assume $n=2$. The black horizontal lines give the 95\%, 85\% and 50\% upper limits on $d_L^\mathrm{GW}$, after restricting our samples to $d_L^\mathrm{GW} > d_L^\mathrm{EM}$. \textit{Bottom:} Allowed Parameter Regions for the transition steepness, $n$, and screening scale $R_c$, for $D=5$ (blue), $D=6$ (green), and $D=7$ (purple), assuming a waveform of the type Equation~\ref{eqn:defmenwaveform}. The vertical black line gives the 2.5\% lower limit for $d_L^{\mathrm{EM}}$. We use the 5\% lower limit for $R_c$ to set these constraints.}
\end{figure}

Figure~\ref{fig:dgwvrc} (top panel) shows the correlation between $d_L^\mathrm{GW}$ and $R_c$ for $D=5$ (blue), $D=6$ (green), and $D=7$ (purple), and for $n=1$ (solid) and $n=2$ (dashed). As can be seen, a steeper transition (i.e. larger value of $n$) allows for theories to have a smaller screening scale; the steeper the transition, the closer the distance must be to the screening scale for a difference in the physics to be noticeable. Increasing numbers of dimensions also allow for smaller screening radii given the same transition steepness; however, the screening radii cannot be much smaller than the minimum EM distance. This is illustrated in the bottom panel of Figure~\ref{fig:twoparam}, where we plot the allowed regions of parameter space within the $n$--$R_c$ plane for $D=5$--$7$. We use the 5\% lower limit for $R_c$, which corresponds to the 95\% upper limit on $d_L^\mathrm{GW}$ after restricting $d_L^\mathrm{GW} > d_L^\mathrm{EM}$, or the 97.5\% upper limit for $d_L^\mathrm{GW}$ (and 2.5\% lower limit for $d_L^\mathrm{EM}$) for the unrestricted samples. For $R_c \gtrsim d_{L, \rm{min}}^{\mathrm{EM}} = (d_{L,\rm{min}}^\mathrm{GW})^{1/\gamma_\mathrm{max}}$ (black, solid line), larger dimensions allow for softer transitions between GR and the higher-dimensional theories. If $R_c \ll d_{L, \rm{min}}^{\mathrm{EM}}$, then these higher dimensional theories are not allowed. As seen in the upper left of Figure~\ref{fig:twoparam}, the minimum screening radius increases with increasing numbers of dimensions.
These results show that theories with extra dimensions that have no screening mechanisms and that affect gravitational propagation at all scales are disfavored by GW170817. In addition, theories with screening mechanisms must have $R_c \gtrsim 20$ Mpc regardless of the transition steepness.

The final modification to GR we consider is theories in which the graviton has a finite lifetime. In such theories, the GW strain scales as Equation~\ref{eqn:decay}, so that setting $d_L = d_L^\mathrm{EM}$, the decay-length is given by:
\begin{equation}
R_g = \frac{d_L^\mathrm{EM}}{\log \left(d_L^\mathrm{GW} /d_L^\mathrm{EM} \right)}.
\end{equation}
Using our posterior samples for $d_L^\mathrm{GW}$ and $d_L^\mathrm{EM} = (d_L^\mathrm{GW})^{1/\gamma}$, and again restricting $\gamma > 1$ to enforce $d_L^\mathrm{GW} > d_L^\mathrm{EM}$, we find a 5\% lower limit for the decay length of the graviton of $R_g > 138$\,Mpc. Since we know that gravitons must travel at the speed of light {\citep{LIGOboth}}, we infer that the lifetime of the graviton can be given as $t = R_g/c > 4.50\times 10^8$\,yr.

We have only considered waveforms that are the same as GR, up to some overall multiplicative factor. It could be possible to evade these constraints by changing the waveforms in other ways. A full analysis of the LVC data using a more general framework \citep{2014PhRvD..89h2001A,Loutrel2014,2015CQGra..32x3001B} would provide more insight into non-GR waveforms.

\begin{figure}
\includegraphics[width=0.5\textwidth]{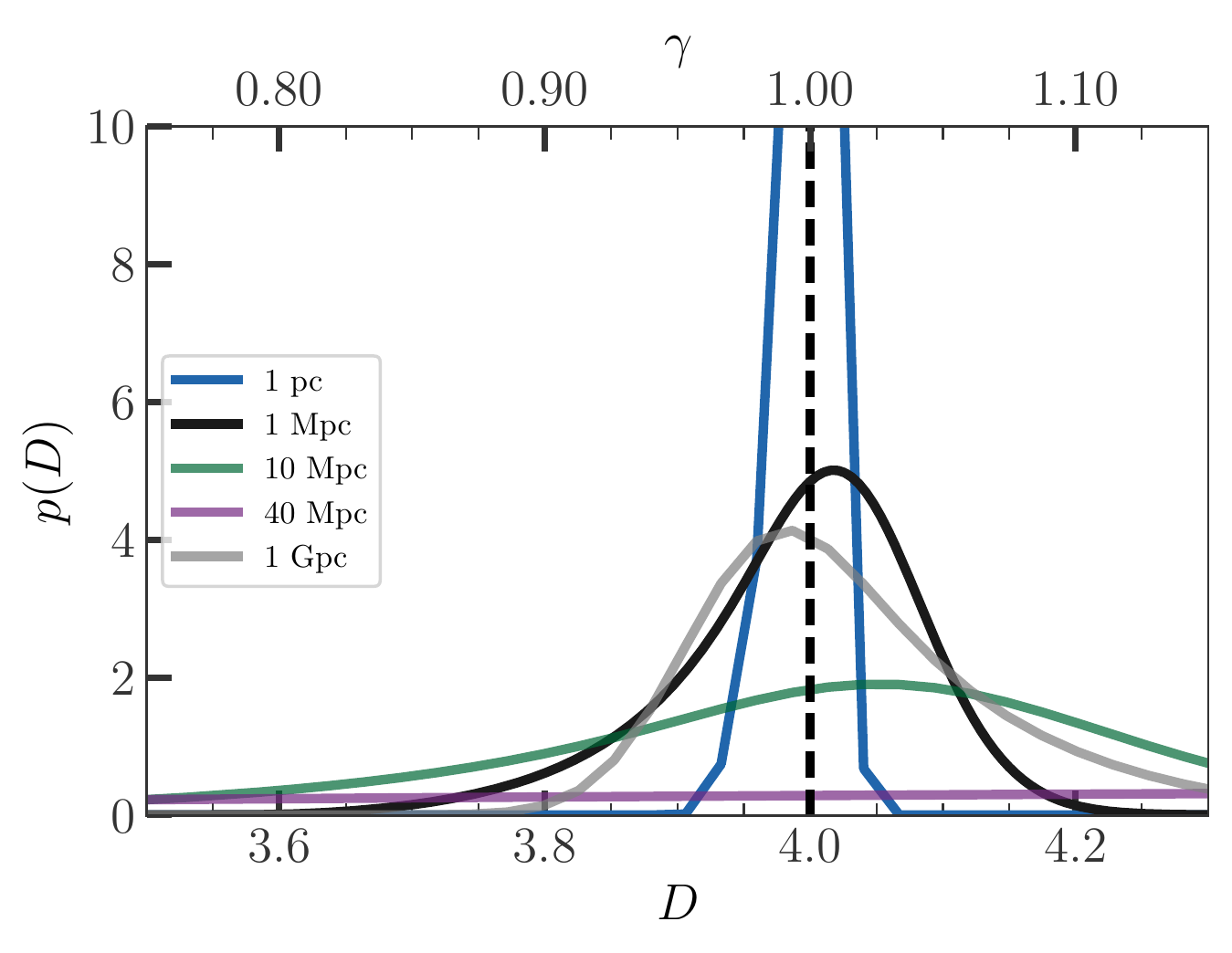}
\caption{\label{fig:diffcrossing} Posterior probability distribution for the number of spacetime dimensions, $D$, assuming different implicit crossing scales. The constraints degrade considerably for a crossing scale equal to the distance to the object, $\sim 40\ \rm{Mpc}$. However, scales either much smaller or larger than this show results that agree well with our choice of crossing scale of 1~{Mpc}.}
\end{figure}

Our analysis relies on a crossing scale for the EM and GW luminosity distances. Equation~\ref{eqn:dampdwaveform} implicitly sets the crossing scale to 1~Mpc, assuming that $h \propto 1/d_L\times (\rm{ 1\ Mpc}/d_L)^{\gamma-1}$. This ensures the correct units for the strain. From a theoretical perspective, the choice of scale is completely arbitrary; our choice of 1~Mpc is motivated by typical galaxy length scales. Figure~\ref{fig:diffcrossing} shows the effects on the posterior for $\gamma$ as a function of different choices for the crossing scale. For scales that are comparable to the distance to GW170817, our constraints degrade considerably, since if the crossing occurs at precisely the distance of the binary then we would be unable to measure deviations as the theory would preclude them by assumption. A crossing scale that happened to be similar to the distance to this particular event would be quite fine-tuned. Scales smaller than a Mpc or larger than a Gpc give similar, or tighter, constraints to what we found above. As we accumulate GW events at different distances, we will be able to fit for the crossing scale directly, in addition to constraining $\gamma$.

We stress that our results do not hold for extra-dimensional theories with compact extra dimensions (e.g. string theory or the ADD model). The extra dimensions need to be at least on the order of the wavelength of the gravitational waves ($\sim 100\ $ km) in order to have a damping effect. In addition, there may be complications for theories with larger extra dimensions.

For example, we find that Randall-Sundrum II and DGP are poorly constrained by GW170817. In Randall-Sundrum II, the massless mode for the graviton is constrained to the 3D-brane; thus, energy cannot efficiently leak into extra non-compact dimension \citep{RS2}. For DGP, only very low frequency waves (i.e. ones with wavelengths on the scale of the cosmic horizon) are allowed to leak into the extra dimension \citep{Dvali2001}.

Our calculation is a phenomenological one---it gives the total damping allowed considering a very general type of leakage for large extra dimensions. Applying these limits to specific theories is beyond the scope of this paper; however, these constraints should be considered carefully by extra-dimensional theories with dimensions of sizes $\sim 100\ \rm{km}$ and greater.

In principle any higher-dimensional theories would allow for extra polarization modes \cite[see, for example,][]{Androit2017}. However, the polarization constraints for GW170817 are quite poor, since the signal was not detected in Virgo and the LIGO detectors are aligned~\citep{LIGO2017}. 
Future events observed by three or more detectors would provide for tighter constraints on extra dimensions.

In this paper we have derived constraints from GW170817 on gravitational leakage by searching for a discrepancy between the measured gravitational luminosity distance, $d_L^\mathrm{GW}$, and the measured EM luminosity distance, $d_L^\mathrm{EM}$. We quantify the gravitational leakage via a damping parameter, $\gamma$, which can be related to the number of non-compact spacetime dimensions, $D$, through which gravity propagates. We find that $D = 4.02^{+0.07}_{-0.10}$ (for SHoES) and $D = 3.98^{+0.07}_{-0.09}$ (for Planck). In addition, we use these constraints to place bounds on extra-dimensional theories with screening mechanisms or decaying gravitons. {We find the graviton decay length to be $R_g>138$\,Mpc, implying a lifetime of the graviton of $t>4.50\times10^8$\,years. In summary, we find that GW170817 is fully consistent with GR.}
\acknowledgements
The authors would like to thank Sylvain Marsat, Richard O'Shaughnessy, Nicola Tamanini, and Ned Wright for helpful comments. KP and MF were supported by the NSF Graduate Research Fellowship Program under grants DGE-1656466 and DGE-1746045. MF and DEH were partially supported by NSF grant PHY-1708081. They were also supported by the Kavli Institute for Cosmological Physics at the University of Chicago through an endowment from the Kavli Foundation. DEH also gratefully acknowledges support from the Marion and Stuart Rice Award.

\appendix
\section*{Statistical Model}
\begin{table*}[!htbp]
\centering
\begin{tabular}{||c|c||c|c||}
\hline 
Variable & Value & Variable & Value \\
\hline
$d_L^\mathrm{GW}$ prior & $\propto (d_L^{GW})^{2}$ & $v_r$  & 3,327 km/s \\
$\gamma$ prior & flat, [0.75, 1.15] & $\sigma_{v_r}, \sigma_{v_p}$ & 72, 239 km/s\\
$H_0$ prior (SHoES) & $\mathcal{N}(\mu_{H_0} = 73.24\ \rm{km/s\ Mpc}^{-1},\ \sigma_{H_0} = 1.74\ \rm{km/s\ Mpc}^{-1})$& $\langle v_p \rangle$ & 310 km/s \\
$H_0$ prior (Planck) & $\mathcal{N}(\mu_{H_0}=67.74\ \rm{km/s\ Mpc}^{-1},\ \sigma_{H_0}=0.46\ \rm{km/s\ Mpc}^{-1})$ & $v_p$ prior & flat, [-1,000,1,000] km/s\\
\hline
\end{tabular}
\caption{\label{tab-methodvalues} Values \& Priors Assumed for the MCMC Analysis} 
\end{table*}

In the following we describe the statistical framework assuming a waveform scaling as in Equation \ref{eqn:dampdwaveform}; however, this is easily extended to any other type of waveform that would cause the GW measurements and EM measurements of the luminosity distance to differ. 

We can write the joint likelihood for the GW data, $x_\mathrm{GW}$, and EM observables, $\langle v_p \rangle$ and $v_r$, given $\gamma$, $H_0$, $d_L^\mathrm{GW}$ and $v_p$ as:
\begin{equation}\label{eqn:jointlikelihood}
\begin{split}
&p(x_\mathrm{GW}, \langle v_p \rangle, v_r \mid \gamma, H_0, d_L^\mathrm{GW}, v_p) \\
&= p(x_\mathrm{GW} \mid d_L^\mathrm{GW})p(\langle v_p \rangle \mid v_p)p(v_r \mid \gamma, H_0, d_L^\mathrm{GW}, v_p),
\end{split}
\end{equation}
where we have assumed that all three observations, $x_\mathrm{GW}$, $\langle v_p \rangle$ and $v_r$ are statistically independent. We can write the third factor in the above equation as:
\begin{equation}
\begin{split}
&p(v_r \mid \gamma, H_0, d_L^\mathrm{GW}, v_p) \\
&= p(v_r \mid v_r^t =v_p+H_0d_L^\mathrm{EM} = v_p + H_0(d_L^\mathrm{GW})^{1/\gamma}),
\end{split}
\end{equation}
where $v_r^t$ is the true recessional velocity of the source.
The likelihoods $p(\langle v_p \rangle \mid v_p)$ and $p(v_r \mid v_r^t)$ are assumed to be Gaussians \citep{H0paper}, and are given as:
\begin{eqnarray}
p(\langle v_p \rangle \mid v_p) &=& \mathcal{N}(v_p, \sigma_{v_p}^2)(\langle v_p \rangle), \\
p(v_r \mid v_r^t) &=& \mathcal{N}(v_r^t, \sigma_{v_r}^2)(v_r).
\end{eqnarray}
Applying Bayes' theorem, the joint posterior for $\gamma$, $H_0$, $d_L^\mathrm{GW}$ and $v_p$ is then:
\begin{widetext}
\begin{align}\label{eqn:jointposterior}
p(\gamma, H_0, d_L^\mathrm{GW}, v_p \mid x_\mathrm{GW}, \langle v_p \rangle, v_r) \propto p(x_\mathrm{GW} \mid d_L^\mathrm{GW})p(\langle v_p \rangle \mid v_p)p(v_r \mid \gamma, H_0, d_L^\mathrm{GW}, v_p)p_0(\gamma, H_0, d_L^\mathrm{GW}, v_p).
\end{align}
\end{widetext}
The posterior for $\gamma$ is found by marginalizing over all other parameters:
\begin{widetext}
\begin{align}
\label{eqn:gammaposterior}
p(\gamma \mid x_\mathrm{GW}, \langle v_p \rangle, v_r)  = \frac{1}{p_\mathrm{det}(\gamma)} \int p(x_\mathrm{GW} \mid d_L^\mathrm{GW})p(\langle v_p \rangle \mid v_p)p(v_r \mid \gamma, H_0, d_L^\mathrm{GW}, v_p)p_0(\gamma, H_0, d_L^\mathrm{GW}, v_p) d H_0 d d_L^\mathrm{GW} d v_p,
\end{align}
\end{widetext}
where $p_\mathrm{det}(\gamma)$ is a normalization term to account for selection effects and ensure that the integral over all detectable datasets integrates to unity. As shown below, this term is negligible for our analysis.

We choose the prior:
\begin{equation} 
p_0(\gamma, H_0, d_L^\mathrm{GW}, v_p) =  p_0(v_p)p_0(d_L^\mathrm{GW})p_0(\gamma)p_0(H_0).
\end{equation} 
This assumes a flat prior for the peculiar velocity, $p_0(v_p) \propto$ constant. For the GW distance, we use the default ``volumetric'' prior used in the LVC analysis, $p_0(d_L^\mathrm{GW}) \propto (d_L^\mathrm{GW})^{2}.$
For the prior on the Hubble constant, $p_0(H_0)$, we take either the SHoES measurement or the Planck measurement. We choose the prior on $\gamma$ to be flat, so the marginal posterior is proportional to the marginal likelihood. Our results are mildly sensitive to these prior choices; for example, taking a flat prior on $d_L^\mathrm{GW}$ shifts the posteriors towards slightly lower values of $\gamma$, so that the MAP and minimal 68\% credible intervals become $1.00^{+0.04}_{-0.06}$ (SHoES $H_0$) and $0.98^{+0.04}_{-0.06}$ (Planck $H_0$) for a flat $d_L^\mathrm{GW}$ prior. (This alternative prior choice also leads to stricter lower limits on the screening scale $R_c$.)
Except for the conservative value of $\sigma_{v_p} = 239$ km s$^{-1}$ from \cite{Guidorzi2017}, all other variable values and priors are the same as those given in \cite{H0paper}. All of our values and priors are given in Table~\ref{tab-methodvalues}.

The normalization term $p_\mathrm{det}(\gamma)$ in Equation~\ref{eqn:gammaposterior} is given by the integral of the marginal likelihood over all detectable datasets \citep{Loredo:2004, MFG:2016}:
\begin{widetext}
\begin{align}
p_\mathrm{det}(\gamma) &= \int_{\rm detectable} p(x_\mathrm{GW}, \langle v_p \rangle, v_r \mid \gamma) dx_\mathrm{GW} d\langle v_p \rangle dv_r \\
&= \int_{\rm detectable} \int p(x_\mathrm{GW} \mid d_L^\mathrm{GW})p(\langle v_p \rangle \mid v_p)p(v_r \mid \gamma, H_0, d_L^\mathrm{GW}, v_p)p_0( v_p) p_0(H_0) p_0(d_L^\mathrm{GW}) d H_0 d d_L^\mathrm{GW} d v_p  dx_\mathrm{GW} d\langle v_p \rangle dv_r.
\end{align}
\end{widetext}
We follow \cite{H0paper} and neglect the EM selection effects. This is justified because the GW horizon for a BNS system during O2 was only 190 Mpc, whereas an EM counterpart would have been observable at distances greater than 400 Mpc. Thus, the integrals over detectable EM datasets, $\langle v_p \rangle$ and $v_r$ integrate to unity. If we neglect the effects of GW redshifting on the detectability of the GW source (which is valid at these low redshifts), the GW selection effects are a function of GW luminosity distance alone. Defining:
\begin{equation}
\int_{\rm detectable \ x_\mathrm{GW} }p(x_\mathrm{GW} \mid d_L^\mathrm{GW})dx_\mathrm{GW} \equiv p_\mathrm{det}(d_L^\mathrm{GW}),
\end{equation}
we have:
\begin{equation}
\label{eqn:pdetgamma}
\begin{split}
&p_\mathrm{det}(\gamma) \\ &= \int p_\mathrm{det}(d_L^\mathrm{GW})p_0( v_p) p_0(H_0) p_0(d_L^\mathrm{GW}) d H_0 d d_L^\mathrm{GW} d v_p.
\end{split}
\end{equation}
The above equation is independent of $\gamma$, and so we can ignore this term in our analysis.
However, if we had chosen to carry out the analysis by setting a prior on the redshift or $v_H$ rather than GW distance, Equation~\ref{eqn:pdetgamma} would have a $\gamma$ dependence in the term $p_\mathrm{det}(d_L^\mathrm{GW} = (\frac{v_H}{H_0})^\gamma)$, which varies significantly over the posterior support for $\gamma$. In this case, $p_\mathrm{det}(\gamma)$ cannot be neglected. 

To compute the posterior for $\gamma$, we sample directly from the joint posterior given by Equation~\ref{eqn:jointposterior} with an MCMC analysis using the python package PyMC3 \citep{PyMC3}.

\end{document}